\documentclass{jpp}
\usepackage{graphicx}
\usepackage{epstopdf, epsfig}

\shorttitle{Review of 0.1 Problem}
\shortauthor{P.~A.~Cassak, Y.-H.~Liu and M.~A.~Shay}

\title{A Review of the 0.1 Reconnection Rate Problem}

\author{P. A. Cassak\aff{1}
  \corresp{\email{Paul.Cassak@mail.wvu.edu}}, Y.-H. Liu\aff{2} \and
  M.~A.~Shay\aff{3}}

\affiliation{\aff{1}Department of Physics and Astronomy, West Virginia
  University, Morgantown, West Virginia 26506, USA \aff{2}Heliophysics
  Science Division, NASA Goddard Space Flight Center, Greenbelt,
  Maryland 20771, USA \aff{3}Department of Physics and Astronomy,
  University of Delaware, Newark, Delaware 19716, USA}

\begin{document}

\maketitle

\begin{abstract}
  A long-standing problem in magnetic reconnection is to explain why
  it tends to proceed at or below a normalized rate of 0.1.  This
  article gives a review of observational and numerical evidence for
  this rate and discusses recent theoretical work addressing this
  problem.  Some remaining open questions are summarized.
\end{abstract}

\section{Introduction}

Magnetic reconnection is a fundamental plasma process that occurs in
many diverse settings including solar physics, magnetospheric physics,
astrophysics, heliophysics, and fusion energy sciences.  It is a key
element in the rapid release of magnetic energy in solar flares,
geomagnetic substorms, astrophysical transients including anomalous
X-ray pulsars and soft gamma repeaters, and the sawtooth crash and
disruptions in magnetically confined fusion devices \citep{Zweibel09}.
It also plays a crucial role in other plasma processes including the
magnetic dynamo and angular momentum transport in accretion disks
around compact astrophysical objects \citep{Zweibel09} and for
dissipation at the end of the turbulent cascade
\citep{Servidio09,Servidio11,Mallet17,Loureiro17,Cerri17}.  The key
ingredient is that a change in topology of the magnetic field allows
for the release of energy stored in the magnetic field into directed
kinetic energy and heat.

Observations and numerical simulations in a wide variety of settings
suggest that the global rate of magnetic reconnection is approximately
0.1 in normalized units.  This means that the electric field pointing
out of the reconnection plane that drives the reconnection is near 0.1
when normalized to a properly defined reconnecting magnetic field and
Alfv\'en speed.  That this is the case has been known for many years,
but developing a theoretical understanding of why has been extremely
challenging.

If we already know the typical rate, why is it important to solve this
problem?  The rate of reconnection is likely related to the efficiency
of particle acceleration and heating during the reconnection process,
which is a very important aspect of identifying reconnection and
studying its effects remotely such as in astrophysical settings where
directly measuring magnetic fields is not possible.  At Earth's
dayside magnetopause, the reconnection rate is directly related to the
efficiency of solar wind-magnetospheric coupling and the rate at which
the global convection pattern occurs, which is a key aspect of space
weather.  In magnetically confined fusion devices, the rate of
reconnection is related to the rate at which material from the core is
ejected during sawteeth.  Thus, knowing what controls the reconnection
rate in various settings is crucial for applications.

This article presents a review of observational, theoretical, and
numerical evidence of the normalized reconnection rate being
approximately 0.1.  Then, recent work addressing why this is the case
is reviewed.  Some open questions are then discussed.  The reader is
also referred to other recent reviews on this topic
\citep{Bhattacharjee04,Zweibel09,Yamada10,Cassak12,Comisso16}.
 
\section{Observational Evidence for the 0.1 Reconnection Rate}
\label{sec:obsevidence}

\begin{figure}
  \centerline{\includegraphics[width=2.2in]{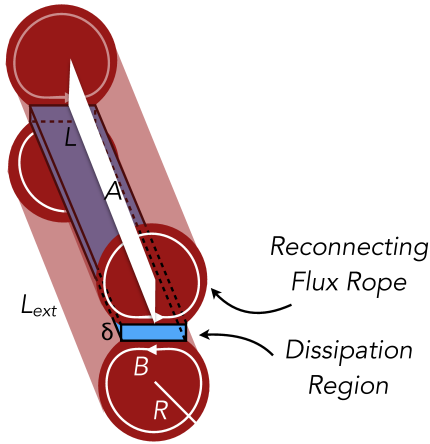}}% Images in 100% size
  \caption{Sketch of two reconnecting flux ropes in red with
    reconnecting magnetic field lines in white.  The flux rope radius
    $R$ and out-of-plane extent $L_{ext}$ are shown.  The dissipation
    region of thickness $\delta$ and length $L$ is in blue.  The white
    surface of area $A$ denotes the location of the magnetic field
    that reconnects in a time $\Delta t$.}
\label{fig:fluxrope}
\end{figure}

The reconnection rate can be inferred with very basic considerations
using plasma parameters observed at macroscopic scales [{\it e.g.,}
  \citep{Shay04}].  Consider two regions of magnetic field coming
together and reconnecting, as shown schematically in
Fig.~\ref{fig:fluxrope}.  Let the time over which significant energy
release via reconnection occurs be $\Delta t$.  Suppose the
reconnecting magnetic field has a characteristic strength $B$
threading a region of characteristic radius $R$ and out-of-plane
extent $L_{ext}$, each assumed uniform for simplicity.  Then, the
magnetic flux processed per unit time by reconnection is $B A /\Delta
t$, where $A \sim R L_{ext}$ as shown by the white surface in the
figure.  From Faraday's law, this must be associated with an electric
field $E$ (the reconnection electric field) extending over a distance
$L_{ext}$ out of the reconnection plane.  In SI units, the relation is
\begin{equation}
  E \sim \frac{BA}{L_{ext} \Delta t} \sim \frac{BR}{\Delta
    t}. \label{eq:egen}
\end{equation}
The reconnection electric field is often presented as a dimensionless
quantity that we call $E^\prime$, which is normalized by the
reconnecting magnetic field $B$ and the Alfv\'en speed $c_A$ based on
$B$ and the ambient plasma density $n$, so
\begin{equation}
E^\prime = \frac{E}{B c_A} \sim \frac{R}{c_A \Delta
  t}. \label{eq:eprimegen}
\end{equation}
There is a geometrical interpretation of this expression.  The
numerator is the radial distance of magnetic flux reconnected in the
time $\Delta t$, and the denominator is the distance that would have
been reconnected in the same time if the inflow speed was $c_A$.  This
is consistent with the more commonly quoted form of $E^\prime =
v_{in}/c_A$, where $v_{in}$ is the inflow speed.

The reconnected rate computed in this way describes the processing of
magnetic flux on a global scale, so we refer to this interchangeably as
the {\it global, large-scale}, or {\it macro-scale} reconnection rate.
However, we point out that the change of magnetic topology takes place
due to dissipation on small scales.  Thus, we distinguish the inferred
macro-scale reconnection rate from the reconnection rate measured
locally near the reconnection site.  This is computed by normalizing
to the magnetic field and plasma parameters immediately upstream of
the dissipation region, so it need not be the same as the global
reconnection rate.  We interchangeably refer to the local rate as the
{\it local, small-scale}, or {\it micro-scale} reconnection rate.  We
are now prepared to estimate the global reconnection rate from
Eq.~(\ref{eq:eprimegen}) for various settings where reconnection
commonly occurs.

\subsection{Solar Flares}

For solar flares, we take plasma parameters from \citet{Priest02}.  In
large solar flares releasing over $10^{24}$ Joules, energy is released
beginning with an impulsive phase that lasts about $\Delta t \simeq$
100 seconds, though energy release continues to a lesser extent over
much longer times.  A magnetic field of strength $B \simeq$ 100 G =
0.01 T threads flux tubes of radius $R \sim 3 \times 10^{7}$ m and
extent $L_{ext} \sim 10^{8}$ m; this provides more than enough
magnetic energy $(B^2 / 2 \mu_0) \upi R^2 L_{ext}$ to power the flare.
Using these parameters, the inferred absolute reconnection rate from
Eq.~(\ref{eq:egen}) is $E \simeq 3000$ V/m, and the normalized global
reconnection rate from Eq.~(\ref{eq:eprimegen}) using $c_A \simeq 4
\times 10^6$ m/s (based on a density of $3 \times 10^{15}$ m$^{-3}$)
is
\begin{equation}
E_{{\rm flare}}^\prime = 0.075. \label{eq:eflare}
\end{equation}

\subsection{Geomagnetic Substorms}

In Earth's magnetotail during geomagnetic substorms, the expansion
phase takes place over a time scale of about 30 minutes
\citep{McPherron70}.  The relevant magnetotail parameters are a lobe
magnetic field $B \simeq 20$ nT, the distance from the plasma sheet to
the magnetopause is $R \simeq 15 \ R_E$, and the cross-tail extent is
$L_{ext} \simeq 30 \ R_E$ \citep{Axford65}, where $R_E$ is the Earth
radius.  The absolute and normalized global reconnection rate
necessary to reconnect the magnetotail in the observed time, using
$c_A \simeq 10^6$ m/s based on a density of 0.1 cm$^{-3}$ = $10^5$
m$^{-3}$ in Eqs.~(\ref{eq:egen}) and (\ref{eq:eprimegen}), is $E
\simeq 1.1$ mV/m (corresponding to a cross-tail potential of 200 kV)
and
\begin{equation}
E_{{\rm substorm}}^\prime \simeq 0.053.
\end{equation}
The similarity of the inferred flare and substorm normalized
reconnection rates prompted \citet{Parker73} to say ``Altogether, the
observations of both solar and magnetospheric activity suggest that
reconnection rates are ``universally'' of the general order of
magnitude of 0.1 $V_A$.''

\subsection{The Sawtooth Crash}

In the sawtooth crash in magnetically confined fusion devices
\citep{vonGoeler74}, the absolute global reconnection rate $E$ should
depend on the size, shape, and plasma parameters in the device in
question.  We are unaware of any comparative studies of normalized
reconnection rates in sawteeth.  However, to get a feel for a
reasonable rate in a modern device, we consider a shot in the Mega
Ampere Spherical Tokamak (MAST) described by \citet{Chapman10}, where
the sawtooth crash time is $\Delta t \simeq 120 \ \mu$s.  The relevant
reconnection parameters were extracted by \citet{Beidler11}: the major
radius is $R_0 = 0.85$ m, so $L_{ext} \simeq 2 \upi R_0$; the $q = 1$
rational surface where reconnection occurs is at $R = 0.32$ m, where
$q$ is the safety factor; the reconnecting (auxiliary) magnetic field
at the upstream edge of the ion diffusion region (at a distance of an
ion Larmor radius upstream of the auxiliary field reversal) is $B
\simeq 6.7$ mT, the corresponding Alfv\'en speed is $c_A \simeq 13$
km/s.  The associated global reconnection rate for processing the
whole core, from Eq.~(\ref{eq:eprimegen}), is
\begin{equation}
E_{{\rm sawtooth}}^\prime \simeq 0.21.
\end{equation}
This value is close to the flare and substorm results.  While a single
shot in a single device does not imply that all devices have the same
rate for all shots, it supports the notion that the normalized
reconnection rate is similar across many settings.

\subsection{More Direct Observations}

Each of these examples provide an {\it indirect} measurement of the
global reconnection rate.  With modern observations, other approaches
to infer the global reconnection rate have become available.  Magnetic
fields at the reconnection site in solar flares cannot be directly
measured, but the speed with which ribbons at the footpoints of two
ribbon flares separate is related to the rate at which magnetic flux
is reconnected higher up in the corona.  Observations give normalized
rates of 0.001--0.2
\citep{Ohyama98,Yokoyama01,Isobe02,Qiu02,Fletcher04,Lin05,Isobe05},
though it should be cautioned that uncertainties in the magnetic
fields to normalize to are quite large.  For the magnetosphere,
reconnection rates can be inferred from the convective electric field;
they are not too different than 0.1 at the dayside
\citep{Mozer10,Wang15} and magnetotail \citep{Blanchard96}, though
again uncertainties are large because there is uncertainty associated
with defining the reconnection plane using techniques such as minimum
variance analysis and its variants.

A direct measurement of the reconnection electric field {\it locally}
has been challenging because of its small size and the same
uncertainties related to finding the appropriate reference frame.  It
has been inferred in laboratory experiments; a value of 0.35 was
reported in the Versatile Toroidal Facility \citep{Egedal07}.
However, using the Magnetospheric Multiscale (MMS) satellites
\citep{Burch16}, the high frequency electric field for an event was
filtered out revealing a DC out-of-plane field in reasonable agreement
with 0.1 \citep{Chen17}.  Thus, there appears to be indirect and
direct observational support that the rate typically is approximately
equal to, and potentially less than, 0.1.

\section{Early Numerical and Theoretical Efforts}
\label{sec:types_paper}

\begin{figure}
  \centerline{\includegraphics[width=3.4in]{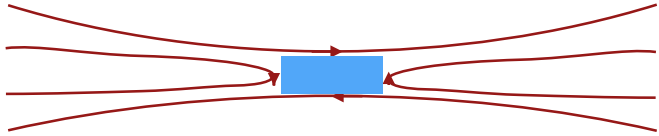}}% Images in 100% size
  \caption{Sketch of reconnecting and reconnected magnetic field
    lines in red, with the dissipation region denoted by the blue rectangle.}
\label{fig:recon}
\end{figure}

Before reconnection was discovered \citep{Dungey53}, there was no
explanation of how magnetic energy could be converted so rapidly.
Magnetic diffusion was orders of magnitude too slow.  Before treating
how reconnection can be fast enough to explain the observed rates, we
point out that theoretical and numerical studies on this problem have
typically been performed using simplified systems.  Unless otherwise
noted, the treatment here is for two-dimensional reconnection with
anti-parallel reconnecting magnetic fields ({\it i.e.,} no ``guide''
magnetic field), as sketched in Fig.~\ref{fig:recon}.  (The presence
of a uniform guide field is not expected to qualitatively or
quantitatively alter the considerations that follow, except as
discussed in Sec.~\ref{sec:questions}.)  We assume the two sides of
the reconnection region are symmetric and isotropic, and the upstream
plasma has no upstream bulk flow and is laminar as opposed to
turbulent.  We also only treat spontaneous reconnection in the
steady-state, meaning the resultant rate from perturbing a current
sheet and waiting until the reconnection reaches a quasi-steady state,
as opposed to continually forcing the reconnection with given upstream
conditions.  The process is studied locally near the dissipation
region, and most studies have not attempted to relate the local
physics to global considerations from far upstream of the reconnection
site.  We make these assumptions not with the idea that all
reconnection satisfies these restrictions, but that the process is
easier to study in this limit and simulations reveal that it contains
sufficient physics to reproduce the 0.1 reconnection rate.

\citet{Parker57} performed a scaling analysis of the reconnection
process providing what we now call Sweet-Parker theory; based on
conservation of mass, energy, and magnetic flux, he showed that the
reconnection rate scales as
\begin{equation}
E \sim \frac{\delta}{L} v_{out} B,  \label{spscale}
\end{equation}
where $\delta$ is the dissipation region thickness in the inflow
direction and $L$ is its length in the outflow direction, $v_{out}$ is
the outflow speed (which scales as the Alfv\'en speed $c_A$), $B$ is
the reconnecting magnetic field strength, and he showed that $\delta /
L \sim 1/S^{1/2}$ when the mechanism allowing the magnetic topology to
change is a uniform resistivity, where $S$ is the Lundquist number.
The {\it local} reconnection rate, normalized to $B$ and $c_A =
v_{out}$, is $E^\prime \sim \delta / L$.  He originally suggested
reconnection could be fast enough to explain flares, but he soon
realized that the prediction was too slow \citep{Parker63}.
Physically, the failure of the Sweet-Parker model is that the exhaust
region closes down into an elongated region.  Since the outflow is
constrained to leave at the Alfv\'en speed, conservation of mass
requires the inflow speed to be small.  Despite the rate being too
small to explain observations, it is still a physically correct model
that accurately describes resistive reconnection with sufficiently
high resistivity.  It has been confirmed in simulations
\citep{Uzdensky00,Cassak05} and experiments
\citep{Ji98,Trintchouk03,Furno05}.  The disagreement with observed
rates spurred research to figure out why reconnection is faster than
what is predicted by the Sweet-Parker model.  We point out in passing
that it was later determined that Sweet-Parker reconnection is
untenable for large enough $S$; instead the Sweet-Parker layer breaks
up producing secondary islands, which fundamentally changes the rate.
Since our treatment here is only for a single X-line, we postpone a
discussion of this important effect to Sec.~\ref{sec:questions}.

The \citet{Petschek64} model predicts local reconnection rates closer
to the observationally inferred global rate.  In the Petschek model,
the curved magnetic fields in the exhaust straighten out, which
produces slow shocks propagating out along the magnetic field that
accelerate the plasma into the outflow jet and provide a more open
exhaust region.  The dissipation region in the Petschek model is more
localized than in the Sweet-Parker model.  The model caused excitement
that the problem was solved, but work much later benefiting from
numerical simulations revealed that the Petschek model is not
self-consistent in the MHD model with a uniform resistivity
\citep{Biskamp86,Uzdensky00}.  However, other models (as described
below) do produce shocks bounding open exhausts
\citep{Liu12,Innocenti15}, so many aspects of the Petschek model are
believed to be essentially correct.  

An advance occurred when it was realized that a fluid model with a
localized (also called ``anomalous'') resistivity could produce
Petschek-like reconnection with the associated high rates
\citep{Ugai77,Sato79}.  Physically, a region of stronger diffusion
near the reconnection site causes the magnetic field line to bend in,
which gives an open outflow exhaust \citep{Kulsrud01}.  There have
been efforts to identify physical causes of anomalous resistivity
[{\it e.g.,} \citep{huba77a,Ugai84,Strauss87}], but it is not well
established that the reconnection rate can be explained through this
mechanism.

An alternate approach considers reconnection in essentially
collisionless systems, for which the scale of collisional diffusion is
much smaller than electron and ion gyroradius scales.  Calculations
\citep{Drake77,Terasawa83,Hassam84} and simulations \citep{Aydemir91}
suggested that the linear phase of collisionless reconnection (the
tearing mode) was much faster than its collisional counterpart.
Numerical studies of steady-state collisionless reconnection in which
the electron to ion mass ratio \citep{Shay98b,Hesse99,Shay07} and
ratio of system size to ion inertial length \citep{shay99a} were
varied found that the local reconnection rate was approximately 0.1
independent of those two quantities.  This prompted the idea that
collisionless reconnection had a ``universal'' rate \citep{shay99a}.
The GEM Challenge study compared simulations with different models,
and found that all models with the Hall term (two-fluid, hybrid, and
full particle-in-cell) had rates comparable to the 0.1 value
\citep{Birn01}.  Many concluded from this that the Hall term was the
cause of reconnection having a rate of 0.1.  While no there is no
first-principles theory showing the reconnection rate is 0.1, it was
suggested that the dispersive behaviour of the Hall term gives rise to
faster flows at smaller scales, which causes the outflow exhausts to
open up as is necessary for reconnection rates faster than
Sweet-Parker \citep{Mandt94,Rogers01,Drake08,Cassak10}.  It should be
pointed out that the Hall term is {\it sufficient} to give rates near
0.1, though it is of course not {\it necessary} (since it was already
known that the same rates arose in MHD with a localized resistivity).

The situation became more complicated when other systems were also
found to have a similar local reconnection rate.  An electron-positron
plasma is potentially important for some astrophysical applications
including pulsar winds and extragalactic jets, but is also important
for fundamental physics because the Hall term in such a plasma is not
present when the electron and ion temperatures are equal
\citep{Bessho05}.  Simulations have revealed that the rate in such a
system is also of order 0.1
\citep{Bessho05,Bessho07,Hesse07,Daughton07,Swisdak08,Zenitani08}.
The mechanism causing the reconnection rate to be close to 0.1 remains
controversial; three different possible mechanisms include secondary
islands \citep{Daughton07}, off-diagonal elements of the pressure
tensor \citep{Bessho07}, and the Weibel instability \citep{Swisdak08}.
When there is a large out-of-plane (guide) magnetic field, the Hall
term can become inactive; reconnection in such a regime also has a
similar rate \citep{Liu14,Stanier15,Cassak15}.  These results confirm
that the Hall term and their associated dispersive waves are not
necessary to get reconnection rates near 0.1.  

The reconnection rate has also been studied in relativistic
reconnection, often with astrophysical applications in mind.  Most
numerical studies have been for electron-positron plasmas because of
the reduced numerical expense.  The global reconnection rate continues
to be close to 0.1 \citep{Bessho12,Liu15,Sironi16}, though
interestingly the local rate can approach 1 \citep{Liu15}.
Relativistic electron-proton reconnection has a similar global rate as
well \citep{Melzani14}.  A similar rate was found in PIC simulations
of reconnection in high energy density laser plasmas \citep{Fox11}.
In summary, the global reconnection rate in a wide variety of systems
described by different physical models and simulated with different
simulation tools has a normalized reconnection rate of approximately
0.1, but it is not clear why this is the case.  We must conclude that
either multiple disparate mechanisms mysteriously give rise to the
same rate, or there is something more fundamental causing the similar
rates.

\section{Recent Insights}

The fact that the (local) reconnection rate is completely altered by
going from a uniform to a localized resistivity within the MHD model
and the similar stark differences between reconnection rates for
systems with and without the Hall term undoubtedly led researchers to
focus on the physics of the dissipation allowing magnetic topology to
change ({\it i.e.,} local physics), to try to solve the 0.1
reconnection rate problem.  However, the evidence discussed in the
previous section suggests this was a misleading approach.  With the
exception of reconnection in MHD with a uniform resistivity (with or
without secondary islands), the {\it global} reconnection rate in all
models tends to be the same.  This suggests that the reconnection rate
is set not by the local (micro-scale) physics allowing the
dissipation, but is due to constraints at the MHD (macro-) scale since
the only aspect all different forms of reconnection have is that they
match up with MHD at the large scales.

Very simple considerations are sufficient to see how large scale
physics could limit the global reconnection rate.  In the Sweet-Parker
model, the local reconnection rate scales like the aspect ratio of the
diffusion region, as shown in Eq.~(\ref{spscale}).  This suggests that
elongated layers are associated with slower reconnection, and opening
up the exhaust angle increases the reconnection rate.  This is
certainly true of Sweet-Parker reconnection, but there is a
limitation.  If one continues to open up the exhaust, one eventually
reaches a system where the exhaust opening angle is 90$^\circ$.
Rather than having a normalized reconnection rate of 1 (with the
inflow speed equal to the Alfv\'en speed) as would be predicted by
$E^\prime = \delta / L$, one actually gets $E^\prime = 0$!  This is
the case because the free energy available for release decreases with
increasing opening angle \citep{Jemella03}, and it vanishes in the
limit of a 90$^\circ$ opening angle (Cassak and Shay, unpublished,
2008).  Equivalently, the symmetry between the inflow and outflow
regions for a 90$^\circ$ opening angle prevents spontaneous
reconnection \citep{Comisso16}.  This implies there is a limitation
due to the large-scale physics that prevents global reconnection rates
of reaching 1.  There must be a maximum global reconnection rate that
is faster than Sweet-Parker reconnection but slower than 1.

\begin{figure}
  \centerline{\includegraphics[width=5.0in]{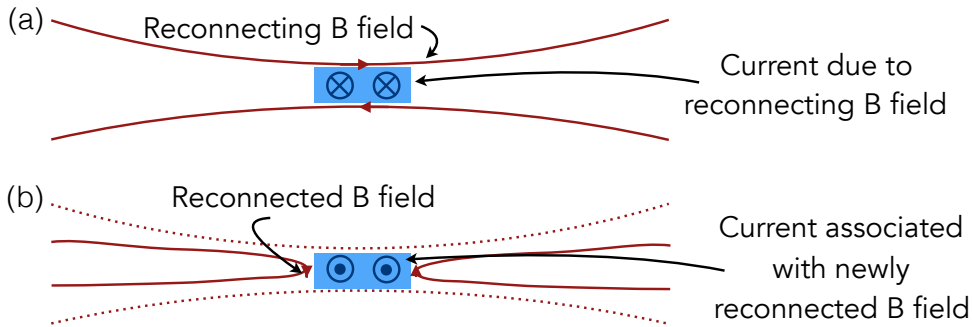}}% Images
                                                             % in 100%
                                                             % size
  \caption{(a) Magnetic field lines in red before they reconnect, with
    their associated current in the dissipation region (in blue) into
    the page.  (b) Magnetic field lines after they reconnect; the
    current due to the newly reconnected field lines opposes the
    current in panel (a).  This motivates why reconnection cannot be
    made arbitrarily fast.}
\label{fig:ka}
\end{figure}

What is the physical reason that the global reconnection rate
decreases when the exhaust opening angle is increased past some
threshold value towards 90$^\circ$?  A qualitative way to understand
this is to consider oppositely directed magnetic fields that are about
to reconnect, as in Fig.~\ref{fig:ka}(a).  The spatial variation in
the upstream (horizontal) field requires an out-of-plane current, in
this case into the page.  Once those magnetic field lines reconnect,
as in Fig.~\ref{fig:ka}(b), reconnected (vertical) fields appear.
From Amp\'ere's law, the current density associated with the newly
reconnected components of the magnetic field is out-of-the-plane, but
it opposes the direction of the background current.  If one were to
make reconnection faster, the reconnected magnetic fields would be
stronger, so the current associated with the reconnected magnetic
fields would become larger.  This weakens the overall current, which
can be seen directly from Amp\'ere's law evaluated at the blue box in
Fig.~\ref{fig:ka} [modified from \citep{Hesse09}]; evaluating the
integral form $\oint {\bf B} \cdot d{\bf l} = \mu_0 \int {\bf J} \cdot
d{\bf A}$, where ${\bf B}$ is the magnetic field, ${\bf J}$ is the
current density, $d{\bf A}$ is the area element directed out of the
page, and $d{\bf l}$ is an infinitesimal path length around the
exterior in a clockwise direction gives
\begin{equation}
J_z = \frac{1}{\mu_0} \left( \frac{B_{x}}{\delta} - \frac{B_y}{L}
\right), \label{jzscale}
\end{equation}
where $B_x$ is the magnitude of the horizontal field at the top and
bottom borders of the box, $B_y$ is the magnitude of the vertical
field at the left and right edges, $J_z$ is the out-of-plane current
density, and $\delta$ and $L$ are the half thickness and length of the
blue box.  In the incompressible limit, which is sufficient for our
purposes, $B_{y}/B_{x} \sim \delta / L$, so Eq.~(\ref{jzscale})
becomes
\begin{equation}
J_z = \frac{B_{x}}{\mu_0 \delta} \left(1 - \frac{\delta^2}{L^2}
\right), \label{jzscale2}
\end{equation}
As $B_y \rightarrow B_x$ ({\it i.e.,} $\delta \rightarrow L$), then
$J_z$ decreases, approaching zero as $B_y$ equals $B_x$.  As the
current decreases, the ${\bf J} \cdot {\bf E}$ work done by the
electric field associated with the dissipation required to change
magnetic topology decreases, and the reconnection slows.  The limit
where the current due to the reconnected field equals the current due
to the reconnecting field would have a reconnection rate of zero,
which is the case of a 90$^\circ$ exhaust opening angle.

As the opening angle increases, the outflow speed decreases
\citep{Hesse09}.  A scaling analysis of the $x$ component of the
momentum equation that balances the convection term $\rho ({\bf v}
\cdot \nabla) v_x$ with the $({\bf J} \times {\bf B})_x = J_z B_y$
(Lorentz) force reveals that the outflow speed $v_x = v_{{\rm out}}$
scales like \citep{Hesse09}
\begin{equation}
v_{{\rm out}} \sim c_{A} \sqrt{1 -
  \frac{\delta^2}{L^2}}. \label{eq:outflow}
\end{equation}
Physically, the reason the outflow speed decreases with increasing
opening angle is because the magnetic pressure term [related to the
  second term in Eq.~(\ref{jzscale})] increases with increasing
$\delta$ and cancels out the magnetic tension force [related to the
  first term].  Since the reconnection rate is proportional to
$v_{out}$, as shown in Eq.~(\ref{spscale}), this suggests the
reconnection rate vanishes for an opening angle of 90$^\circ$.
However, using this expression in Eq.~(\ref{spscale}) predicts a
maximum rate near 0.4, which is noticeably larger than the
reconnection rate obtained in simulations or observations.

A potentially important development came recently, when it was argued
that there is a second reason that the reconnection rate decreases
when increasing the opening angle \citep{Liu17}.  When the opening
angle is increased, the thickness of the current sheet can exceed the
length scales associated with the micro-scale processes that allow
magnetic topology to change.  This happens because the micro-scale
physics typically have a fixed scale which defines the region over
which they are important.  For example, collisionless anti-parallel
reconnection in an electron-proton plasma has micro-scales set by the
ion and electron inertial scales $d_i = c / \omega_{pi}$ and $d_e = c
/ \omega_{pe}$ \citep{Sonnerup79}.  The dissipation region, therefore,
cannot respond by getting arbitrarily thick.  Instead, the system
responds by developing a three-scale structure with the micro-scale
set by dissipation physics, the macro-scale set externally, and an
intermediate meso-scale that joins the two.  The micro-scale
dissipation region becomes embedded within the meso-scale structure.

\begin{figure}
  \centerline{\includegraphics[width=4.0in]{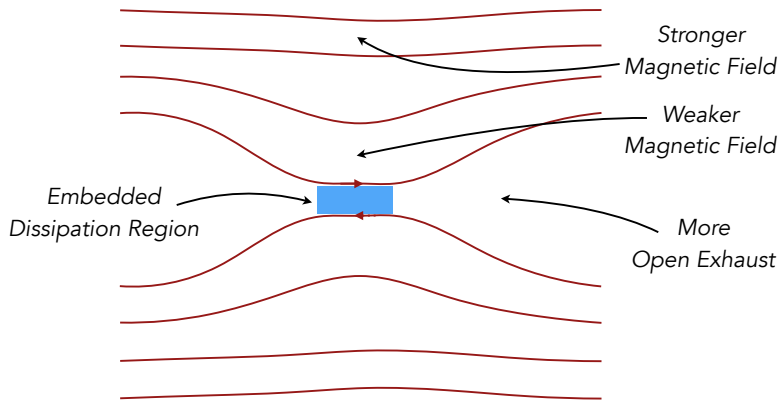}}% Images in 100% size
\caption{Sketch of the reconnection region if the opening angle of the
  exhaust were to be made more open.  This sketch motivates that the
  magnetic field at the dissipation region becomes weaker because the
  dissipation region remains at micro-scales, after \citet{Liu17}.}
\label{fig:yihsin}
\end{figure}

This has important consequences.  For systems with low plasma $\beta$,
the reconnecting magnetic field immediately upstream of the
dissipation region gets weaker, as is sketched in
Fig.~\ref{fig:yihsin}.  Physically, this occurs because the curvature
force away from the current sheet in the upstream region must be
nearly balanced by the magnetic pressure gradient force towards the
current sheet.  This implies the reconnecting magnetic field at the
reconnection site is weaker than the large-scale magnetic field, so
the local magnetic field $B$ driving the outflow is weaker.  This
decreases the reconnection rate, as seen in Eq.~(\ref{spscale}), both
because the reconnection rate is directly proportional to $B$ and
because the Alfv\'en speed is directly proportional to $B$.  A short
calculation on very general grounds showed that the local magnetic
field $B$ is
\begin{equation}
B \sim B_{0} \left(\frac{1 - \delta^2 / L^2}{1 + \delta^2 / L^2}
\right),
\end{equation}
where $B_0$ is the asymptotic macro-scale magnetic field.  When
convolved with the outflow speed in Eq.~(\ref{eq:outflow}) to find the
reconnection rate in Eq.~(\ref{spscale}), it is found that the maximum
in the normalized {\it global} reconnection rate is $E^\prime \simeq
0.2$ \citep{Liu17}, much closer to the observed maximum rate.  It was
further shown that the weakening of the upstream reconnecting field is
more important to the scaling than the weakening of the outflow speed.
The results compared favorably with particle-in-cell (PIC) simulations
of both non-relativistic electron-proton and relativistic
electron-positron plasmas \citep{Liu17}.  This result presented a new
motivation for why the reconnection rate is close to, and bounded
above by, 0.1.

In summary, these new insights suggest that the constraints of energy
release at the meso- and macro-scale play a crucial role in setting
the maximum global reconnection rate.  An appealing aspect of this
approach is that it provides a natural explanation for why the global
reconnection rate is the same in the varied numerical systems where a
rate of 0.1 has been obtained with vastly different micro-scale
physics.  It also properly implies that the reconnection rate can only
be between 0 and 1, which is not the case in the \citet{Petschek64}
model.

\section{Some Open Questions}
\label{sec:questions}

The previous section does not imply the 0.1 problem is solved.  We
outline a number of open questions:
\begin{itemize}
  \item Perhaps the most important unsolved problem is, even
    continuing to make the simplifying assumptions made here, why is
    the global reconnection rate not always the maximum rate near 0.1?
    There is a distinct difference between collisionless reconnection
    and reconnection in MHD with anomalous resistivity on the one
    hand, and reconnection in MHD with a uniform resistivity on the
    other.  For the latter, if the resistivity is above a threshold
    [where the Lundquist number $S$ based on the length of the current
      sheet is below $10^4$ \citep{Biskamp86,Loureiro07}], collisional
    reconnection proceeds as outlined by the Sweet-Parker model rather
    than with a reconnection rate of 0.1.  If the resistivity is
    uniform but below the same threshold, secondary islands
    spontaneously arise and the normalized reconnection rate is
    \citep{Bhattacharjee09,Cassak09a,Huang10}
\begin{equation}
  E^\prime \simeq 0.01. 
\end{equation}
It is important to point out that there is a distinct difference
between the 0.01 rate when secondary islands are present in resistive
reconnection and the 0.1 rate of collisionless reconnection
\citep{Daughton09,Shepherd10}.  Why does Sweet-Parker reconnection not
proceed at the maximum rate?  When secondary islands arise and make
reconnection faster, why is the reconnection rate limited by 0.01
instead of 0.1?  Answering both of these questions is crucial to a
full understanding of the 0.1 problem.

\item A related question is the following: is the resistive-MHD
  description of magnetic reconnection ever valid other than
  potentially transiently?  Since the observed reconnection rates are
  close to 0.1, does that imply that the slower modes of Sweet-Parker
  reconnection and resistive reconnection with secondary islands are
  not the dominant mode in the release of magnetic energy by
  reconnection?  It has been suggested that resistive reconnection
  could potentially occur at the early phase of the energy release,
  but that a transition to faster reconnection is necessary for the
  rapid energy release \citep{Shibata01,Cassak05,Uzdensky07} and it
  has been pointed out that reconnection electric fields are
  significantly stronger than the Dreicer electric field so that
  classical resistivity is not possible [{\it e.g.},
    \citep{Cassak13}], while others argue that resistive effects are
  sufficient and can explain the energy release.
   
\item How does the reconnection rate change when we start relaxing our
  assumptions?  When the upstream plasmas are asymmetric, the
  reconnection rate remains 0.1 \citep{Cassak08b,Malakit10} when
  properly normalized \citep{Cassak07d}.  However, diamagnetic effects
  in the presence of a guide field \citep{Swisdak03,Liu16} or the
  presence of an upstream flow \citep{cassak11a,Doss15} can reduce the
  reconnection rate and even prevent reconnection all together.  The
  effect of the diamagnetic drift is interesting because it is a
  finite Larmor radius effect that is not present in the MHD model.
  Thus, local finite Larmor radius effects can also impact the global
  reconnection rate, which along with the first bullet point is
  another example that the global reconnection rate is not purely set
  by macro-scale physics like the model in the previous section might
  suggest.  How does the slowing of reconnection due to diamagnetic
  effects fit in to the picture of maximizing the global rate of
  reconnection?

\item How does the local reconnection picture described in the
  simplified models fit into global large-scale physics in the context
  of applications?  This is an important problem even in more
  complicated idealized reconnection studies than the ones discussed
  here.  In the island coalescence problem, two like current channels
  attract each other, causing reconnection in the magnetic fields
  between them.  If the current channels are much bigger than kinetic
  scales, the flux ropes can bounce off each other
  \citep{Karimabadi11b,Stanier15b,Ng15} because the time scale for
  reconnection can become longer than the time scale of the
  interaction.  The global reconnection rate in that case is quite
  small, {\it i.e.,} reconnection does not always proceed at the 0.1
  rate discussed here.  However, if the local reconnection rate in the
  embedded current sheet \citep{Shay04} is measured, the local
  reconnection rate is 0.1 \citep{Karimabadi11b}.  Such questions are
  crucial for applications of reconnection to all settings, including
  solar, magnetospheric, astrophysics, and fusion.  For example, it is
  particularly important for flux ropes in the corona which are far
  larger than kinetic scales.  How can a theory be developed to bridge
  the macro- and micro-scales?  The approaches by \citet{Simakov06}
  and \citet{Liu17} might be useful.

\item Why does reconnection not always proceed, {\it i.e.,} what
  causes reconnection onset?  For example, in the corona before a
  flare and the magnetotail before a substorm, current sheets must
  exist because they are associated with free energy, but large-scale
  reconnection does not occur.  Large amounts of energy could not be
  stored if the current sheets were to always reconnect with a rate of
  0.1 \citep{Dahlburg05} or even 0.01 \citep{Cassak09b}.  What
  prevents the release of energy at the maximum allowable rate during
  times when energy is being stored?

\item Why do results based on idealized two-dimensional systems seem
  to work so well in producing a rate that is consistent with the
  rates inferred from global observations which are three-dimensional?
  Is reconnection really typically quasi-2D?  There are many examples
  of successful direct comparisons between magnetospheric satellite
  data and 2D numerical simulations, which suggests reconnection may
  be a quasi-2D process.  More recently, MMS satellites have observed
  strong electric fields that arise in 3D simulations but not in 2D
  [{\it e.g.}, \citep{Burch16b,Ergun16}], so clearly 3D effects can be
  important.  However, when quasi-2D reconnection is averaged over the
  out-of-plane direction even in a system with complicated
  reconnection patterns, the same reconnection rate near 0.1 appears
  (W. Daughton, plenary talk at 2016 US-Japan Meeting on Magnetic
  Reconnection), so it is not clear how the observed 3D structure
  impacts the global reconnection rate.  Can the global reconnection
  rate in 3D systems also be explained as a result of constraints on
  the global energy release or does it require more general approaches
  \citep{Pontin11,Boozer13}?  More fundamentally, what conditions must
  be met for the 2D model to be accurate?  In one study, reconnection
  with current sheets of small extent in the out-of-plane direction
  had lower reconnection rates than 2D reconnection, suggesting that
  reconnection is more energetically favorable when it is quasi-2D
  than fully 3D \citep{Meyer13}.  This was attributed to the ends of
  the X-line being energy sinks.  Significantly more work is needed on
  this problem.  Another interesting open question is: how does
  upstream turbulence \citep{Matthaeus85,Smith04} impact the global
  reconnection rate?
\end{itemize}

The authors acknowledge wonderful conversations on this problem with
many community members over the years.  We acknowledge support from
NSF Grant AGS-1460037 (PAC) and NASA Grants NNX16AG76G (PAC),
NNX16AG75G (YL) and NNX15AW58G (MAS).  This manuscript was prepared in
advance of the first Journal of Plasma Physics Frontiers in Plasma
Physics Conference in anticipation of lively conversations.

%\bibliographystyle{jpp}
%\bibliography{gcrbib}
%%\bibliography{jpp-instructions}

\end{document}